\documentclass[a4paper,10pt]{edpsci}

\usepackage{amsmath}
\usepackage{amssymb}
\usepackage{graphicx}

\usepackage[naustrian,english]{babel}
\usepackage[T1]{fontenc}
\usepackage[utf8]{inputenc}
\usepackage{csquotes}
\usepackage{xcolor}

\usepackage{hyperref}
\hypersetup{
    colorlinks=true,
    citecolor=black,
    urlcolor=black,
    linkcolor=black
}

\usepackage[%
    backend=biber,%
    style=numeric,%
    doi=true,%
    url=false,%
    isbn=false,%
    sorting=none,%
    maxcitenames=2, 
    maxbibnames=12,
    minbibnames=3
]{biblatex}
\journalname{Acta Acustica}
\articletype{Research article}

\usepackage{enumitem}
\usepackage{booktabs}
\usepackage{multirow}
\usepackage{comment}

\usepackage{xcolor}

\definecolor{rankgood}{RGB}{225,245,230}

\newcommand{\rankval}[2]{%
  \ifnum#1=#2%
    \colorbox{rankgood}{#1/#2}%
  \else%
    #1/#2%
  \fi%
}

\begin{document}

\title{Zero-group-velocity modes in textured steel plates: experiments and elastic modeling}

\author{
    C.\ Grünsteidl\inst{1}\correspondingauthor{\email{clemens.gruensteidl@recendt.at}}
    \and
    C.\ Kerschbaummayr\inst{1}
    \and
    G.\ Yan\inst{1}
    \and
    G.\ Watzl\inst{1}
    \and
    F.\ Rudinger\inst{1}
    \and
    E.\ Scherleitner\inst{1}
    \and
    M.\ Sonnleitner\inst{2}
    \and
    C.\ Hoflehner\inst{2}
}

\authorrunning{C.\ Grünsteidl et al.}

\institute{
    Research Center for Non Destructive Testing (RECENDT) GmbH,
    Altenberger Straße 69, 4040 Linz, Austria
    \and
    voestalpine Stahl GmbH,
    voestalpine-Straße 3, 4020 Linz, Austria
}

\abstract{
Zero-group-velocity (ZGV) resonances are investigated as a non-contact ultrasonic probe of elastic anisotropy in textured steel plates. Directional laser-ultrasonic measurements were performed on 70\% cold-rolled Ti-stabilized interstitial-free steel plates in the full-hard and recrystallized condition. Experimental results were compared with predictions based on effective elastic stiffness tensors derived from X-ray diffraction measurements. Classical Voigt--Reuss--Hill estimates and microstructure-resolved finite-element homogenization reproduce the measured angular dependence of the ZGV frequencies well for both states, with the finite-element KUBC/SUBC bounds considerably narrower than the classical Voigt--Reuss bounds. 
For the full-hard state, agreement with the texture-based model is obtained only after a recovery treatment. This indicates that dislocations present in the full-hard material contribute to its elastic response, an interpretation that is further supported by theoretical estimations. An identifiability analysis shows that the directional ZGV measurements provide access to at least three independently observable combinations of the stiffness tensor components, and concluding suggestions for application in recrystallization monitoring are provided.
}


\maketitle

\section{Introduction}
\subsection{Context and motivation}
The elastic properties of engineering metals are strongly influenced by their polycrystalline microstructure. In addition to the single-crystal elastic properties, the crystallographic orientations and spatial morphology of the
grains affect the macroscopic response \cite{Diehl2019}. Forming processes
such as rolling commonly generate preferred crystallographic textures rather than uniform orientation distributions \cite{Daz-Fuentes2004, Ashby1999, Inagaki1994, Kestens2016}. The collective response of many grains can then be described by effective macroscopic material properties that account for texture-induced anisotropy. Within the framework of linear elasticity, this behavior is represented by an effective stiffness tensor, which may be estimated from the single-crystal stiffness tensor and the orientation distribution function (ODF).

In the production of steel sheet, control of the crystallographic texture is of considerable technological importance, as it strongly influences the material's formability and, consequently, its suitability for subsequent manufacturing operations such as deep drawing. During recrystallization of cold-rolled steel sheets, the elongated grains generated by rolling are replaced by a new population of approximately equiaxed grains. This microstructural transformation is typically accompanied by a significant evolution of the crystallographic texture, ultimately determining the effective elastic and plastic properties of the final product \cite{Daz-Fuentes2004, Rollett2017}.

The present work investigates advanced ultrasonic methods for probing the effective elastic behavior of textured sheet metals, with the long-term objective of providing process feedback during recrystallization. Unlike conventional ultrasonic techniques, which often reduce the elastic response to one or a few wave velocities, guided-wave and zero-group-velocity (ZGV) measurements provide a rich set of observables that are sensitive to elastic anisotropy. As a first step, we focus on the fundamental relationship between crystallographic texture and these ultrasonic observables. Experimental results are compared with predictions based on ODF-derived effective stiffness tensors, allowing the sensitivity of the measured resonances and dispersion characteristics to texture-induced anisotropy to be assessed. 

Because guided-wave dispersion and ZGV resonances depend on the complete effective elastic tensor, they also provide a demanding benchmark for the homogenization models.
Beyond assessing texture-induced anisotropy, the comparison may reveal to what extent the material behavior is adequately described by the homogenized elastic tensor alone. Deviations between measured and predicted resonances could indicate the influence of additional factors that are not explicitly included in the model, such as residual stresses, dislocation density, or other aspects of the microstructure.

\subsection{Homogenized elastic properties of textured polycrystals}

To relate crystallographic texture to macroscopic elastic behavior, the effective stiffness tensor of the polycrystal must be estimated from the elastic properties of the constituent single crystals and their orientation distribution.

Classical mean-field estimates are provided by the Voigt and Reuss models, which assume uniform strain and uniform stress throughout the aggregate, respectively. In the Voigt approach, the stiffness tensor of each grain is first rotated into the sample coordinate system according to its crystallographic orientation, and the effective stiffness is obtained by averaging these rotated stiffness tensors over the orientation distribution. In contrast, the Reuss model averages the correspondingly rotated compliance tensors and subsequently inverts the result to obtain the effective stiffness tensor. These assumptions generally yield upper and lower energetic estimates of the effective stiffness, whereas the Voigt--Reuss--Hill approximation uses their arithmetic mean as a practical intermediate estimate. More advanced self-consistent schemes account approximately for elastic interactions between differently oriented grains by embedding each grain in an effective medium whose properties are determined iteratively. These analytical approaches provide efficient estimates of the homogenized elastic tensor, but they do not explicitly resolve the spatial grain arrangement or local intragranular fields. \cite{Hirsekorn1990, R_Hill_1952}

Microstructure-resolved finite-element (FE) homogenization \cite{Kamaya2009,Miehe2002} avoids the uniform-field assumptions by solving the heterogeneous elastic boundary-value problem directly. For a finite statistical volume element, the apparent elastic response depends on the applied boundary conditions. Kinematically uniform boundary conditions (KUBC) prescribe an affine displacement field at the boundary and produce the stiffer apparent response, whereas statically uniform boundary conditions (SUBC) prescribe uniform macroscopic tractions and produce the more compliant response. Their difference decreases as the considered volume becomes increasingly representative of the material \cite{Huet1990}.
While it is also possible to simulate a dynamic response with FE in synthetic polycrystals \cite{VanPamel2015}, we restrict this report on static models, as good agreement with experiments was found.

\subsection{Ultrasonic measurements of anisotropic elastic parameters}

Ultrasonic techniques are well established for determining anisotropic elastic stiffness tensors when measurements can be performed along several propagation and polarization directions. Dever \cite{Dever1972} determined the three independent elastic constants of single-crystalline $\alpha$-iron  in a broad temperature range.  
Measurements and theoretical Voigt--Reuss--Hill-averaged values of the stiffness tensor for textured low-carbon steel were reported by Spalthoff et al. \cite{Spalthoff1993}. Such approaches require access to suitable acoustic propagation paths and may involve specially oriented specimens, multiple specimen cuts, or several measurement configurations. These requirements limit their applicability in quality-control settings, where characterization may need to be performed non-destructively on a single component.

Guided-wave techniques provide an alternative because multimodal wave propagation can be observed along different in-plane directions of a waveguide, such as a plate. Bochud et al., for example, demonstrated the recovery of anisotropic plate properties from directional measurements of multimode guided waves on a single specimen \cite{Bochud2018}.

The experimental constraints become even more severe for in-situ
characterization during thermal processing. In this context, Laser ultrasonics (LUS) is
attractive because it may be applied remotely without mechanical contact with the heated specimen. 
Previous studies have monitored recrystallization and texture evolution using the longitudinal
sound velocity measured in the through-thickness direction \cite{Hutchinson2008,Militzer2013}. 
Although such measurements permit high temporal sampling of the material evolution, a single through-thickness velocity provides only limited information about the full anisotropic elastic
response. Directional mapping can provide additional information and can be implemented using LUS \cite{Bate2017}; however, the required scanning schemes reduce the attainable temporal resolution.

The combination of LUS with guided wave modes, and especially zero-group-velocity (ZGV) plate resonances \cite{Prada2008ZGV} may offer an attractive compromise. 
It provides both experimental simplicity and extended directional information. The suitability of ZGV modes for in-situ laser-ultrasonic measurements during thermal cycles has already been demonstrated \cite{Watzl2022}. 

\subsection{Zero group velocity guided wave modes in anisotropic plates}

ZGV modes \cite{Prada2008ZGV} correspond to local minima in the wavenumber-frequency ($k$-$f$) dispersion curves. 
They are most extensively studied in plates, although they are also known to arise in other waveguide geometries \cite{Ryzy2026Balls}. 
At these points, the energy transport velocity, given by the group velocity $\mathrm{d}f/\mathrm{d}k$ of the guided wave, vanishes. As a result, elastic energy coupled into a ZGV mode remains spatially confined, leading to a pronounced local resonance. 
This manifests as a sustained ringing response at the ZGV frequency. Consequently, ZGV modes appear as distinct spectral peaks in the local response to an appropriately chosen excitation.

LUS is particularly well suited for the investigation of ZGV modes. These modes generally involve both in-plane and out-of-plane displacement components, enabling efficient excitation through a thermoelastic laser source while maintaining high sensitivity to interferometric detection. 
Importantly, the non-contact nature of LUS preserves the intrinsic waveguide characteristics. 
This is essential for the study of ZGV modes, which emerge from the dispersion properties of the unperturbed structure. 
Any attached transducer would locally modify the boundary conditions and waveguide geometry, thereby changing the modal spectrum and potentially destroying the conditions required for the existence of the ZGV mode.

The ZGV frequency is determined by the plate's elastic properties, thickness $h$, and mass density. Owing to this dependence, the frequencies of ZGV resonance peaks have been exploited for the characterization of elastic material properties \cite{Clorennec2007_ZGVSoundVelocity}. 
More recently, approaches combining ZGV resonances with additional guided-wave modes have been proposed to extend the range of accessible waveguide parameters \cite{Watzl2025CoMLAS, Diboune2026SSRN}.

To describe guided waves and ZGV modes in plates composed of anisotropic material, such as single crystal wafers, fibre-reenforced composites or textured metals, the elastic tensor of the material needs to be considered. While this complicates the analytic derivation of dispersion curves \cite{Li1990}, it is straight-forward with available numerical tools \cite{kiefer_GEWTool, Huber2024DispersionCalculator, Pavlakovic1997}. 

\begin{figure}
    \centering
    \includegraphics[width=1\linewidth]{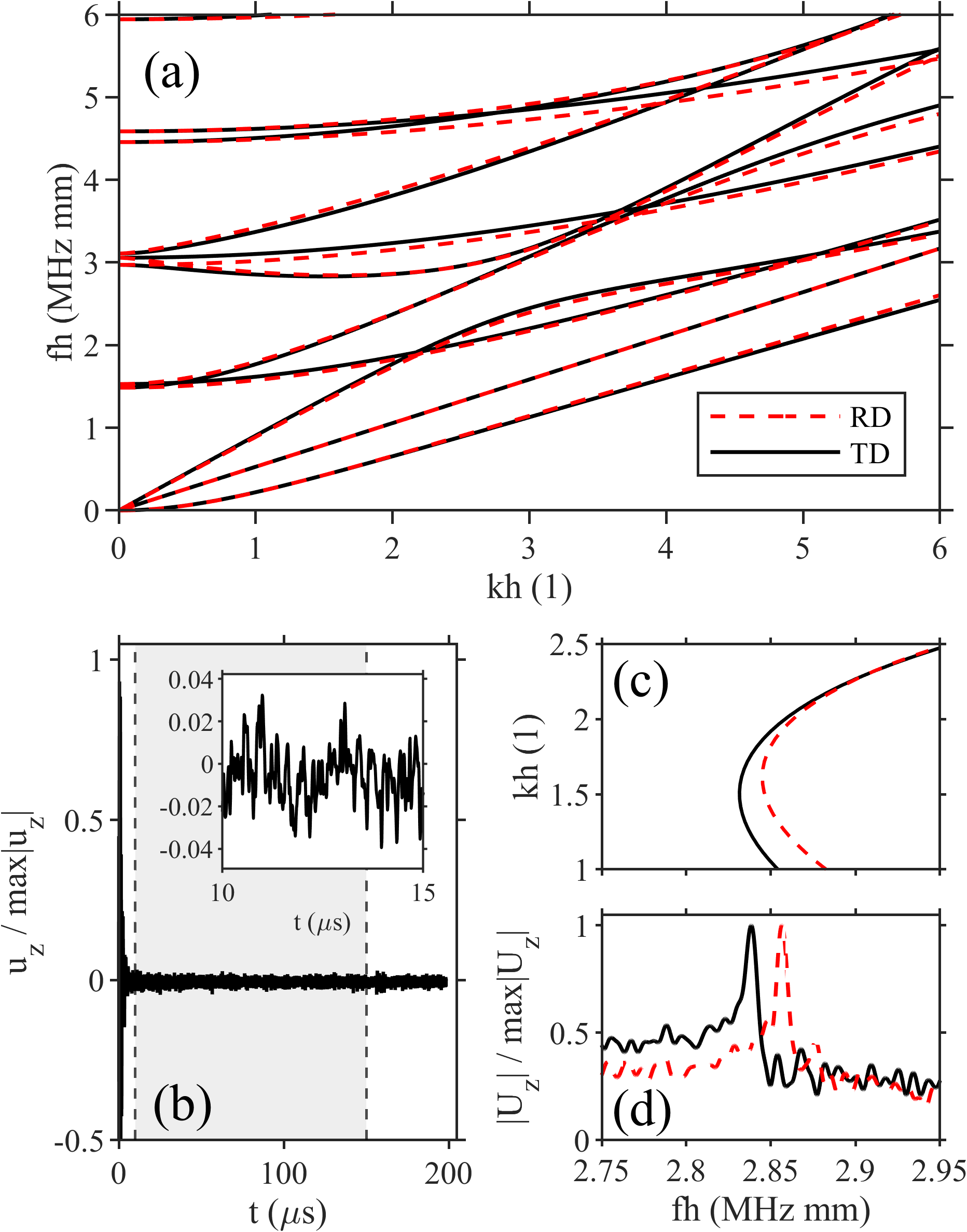}
    \caption{Guided waves and ZGV modes in anistropic steel sample of recrystallized texture; Propagation in rolling direction (RD) and transverse direction (TD) shon; (a) Dispersion curves; (b) Time signal of local LUS response measurement, TD shown only; (c) Zoom into region of lowest order ZGV or dispersion curves; (d) Measured ZGV spectra in RD and TD.}
    \label{fig:disp_time_spec}
\end{figure}

As a consequence of material anisotropy, the dispersion curves become direction-dependent. Away from the principal material directions, different wave polarizations are coupled and can no longer be classified into distinct mode families. 
Figure \ref{fig:disp_time_spec}\,(a) shows thickness-normalized dispersion curves for a cold-rolled and recrystallized steel plate, calculated in rolling direction (RD) and transverse direction (TD) using GEWTool \cite{kiefer_GEWTool}. 
These are the fully coupled solutions, including also shear-horizontal modes. In Figure \ref{fig:disp_time_spec}\,(c) a zoom on the lowest order ZGV modes is provided. 
The resonance peaks in (d) measured in corresponding directions appear at the predicted local minima in $k$. 

In addition, the group-velocity vector is generally not collinear with the wave vector, leading to wave skewing, where energy propagates in a direction different from that of the wavefront normal.

With a focus on ZGV modes, these anisotropy-induced effects have recently been investigated by Kiefer et al. \cite{Kiefer2023}. Already earlier, Prada et al. \cite{Prada2009} demonstrated the directional dependence of ZGV frequencies in a monocrystalline silicon wafer. Using a line-source excitation, guided waves are generated with wave vectors oriented perpendicular to the source line, allowing the angular dependence of the ZGV resonance to be resolved by rotating the specimen.
A point-source excitation launches guided waves in all in-plane directions simultaneously. The measured resonance therefore represents a superposition of contributions from different propagation angles. Rather than yielding a single resonance frequency, the response consists of a broad resonance feature with distinct sub-peaks. As later shown by Kiefer et al. \cite{Kiefer2023}, the sub-peaks correspond to propagation directions for which the group velocity vanishes completely. The response between them is associated with transverse-group-velocity modes, for which the group velocity component parallel to the wave vector vanishes, while a finite transverse component remains.

\section{Experiments}
\subsection{Laser-ultrasonic measurements}
\begin{figure}
    \centering
    \includegraphics[width=1\linewidth]{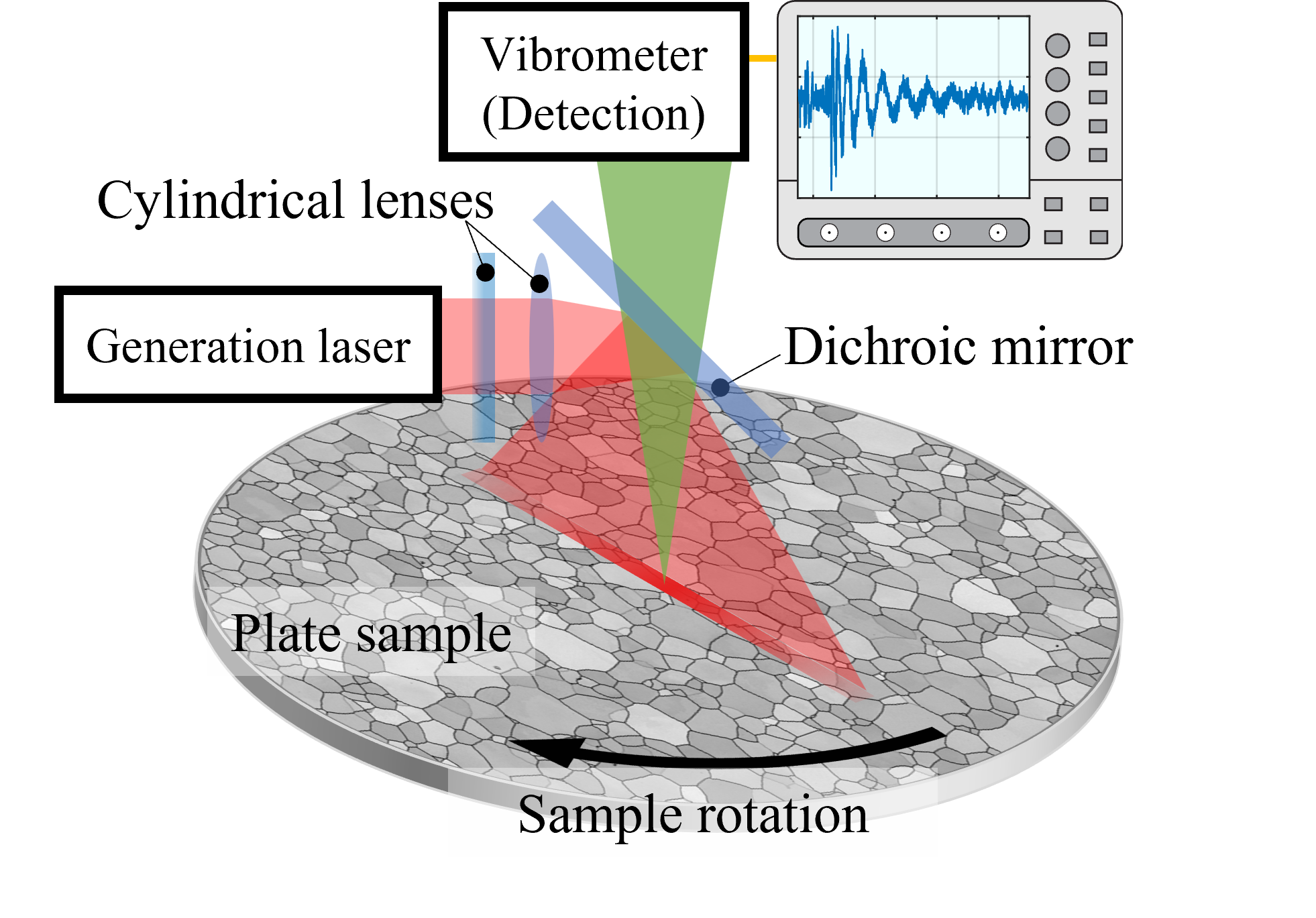}
    \caption{Schematic LUS setup for directional ZGV measurements}
        \label{fig:setup}
\end{figure}

The experimental setup is shown in Fig.~\ref{fig:setup}. Guided waves were generated using a pulsed laser source (Bright Solutions Wedge HB 1064, wavelength 1064\,nm, pulse duration approximately 1.5\,ns, energy 2\,mJ). The beam was focused onto the sample surface using a cylindrical lens with a focal length of 130\,mm, forming a line source. To increase the line length while maintaining a narrow line width, a second cylindrical lens oriented orthogonally to the first was employed.\footnote{In the present implementation, the orthogonal cylindrical lens was realized using a spatial light modulator already integrated into the setup, providing additional flexibility while serving the same optical function.}

Out-of-plane surface displacements were measured at the center of the excitation line using a Sound\&Bright TEMPO two-wave-mixing interferometer with the bandwidth limited to 0.5 - 20\,MHz.

Samples were provided as circular discs with a diameter of 62 mm and a thickness of approximately 1 mm. For the measurement they were mounted on an automated rotation stage. A strip of reusable adhesive putty (UHU Patafix) was applied along the circumference of the specimens. Besides securing the samples on the rotation stage, it provided damping of waves reflected from the plate boundary. The line-source geometry preferentially excites guided waves propagating normal to the source line. Rotating the specimen therefore allows the angular dependence of the guided-wave response to be measured. Angular scans were carried out from $0^\circ$ to $358^\circ$ in steps of $2^\circ$, where for $0^\circ$ the generation line was in parallel with the rolling direction (RD) (thus at $90^\circ$ waves are propagating in RD). For each angular position, signals were averaged over 1000 laser shots. The excitation laser was operated at a repetition rate of 500\,Hz, thus 2\,s per angle position were required.
An example signal and typical ZGV spectra are shown in Figure\ref{fig:disp_time_spec}\,(b) and \ref{fig:disp_time_spec}\,(d). In \ref{fig:disp_time_spec}\,(b) the temporal range used to calculate ZGV spectra is highlighted. A range after early transient signal contributions and remaining edge reflections was chosen to obtain a mostly isolated ZGV peak.

In addition to directional ZGV mode measurements, we also measured the through-thickness longitudinal sound velocity of the samples. The procedure and setup for this is described in \cite{Watzl2022}. Results of all experimental and numerically calculated longitudinal sound velocities are summarized in Fig. \ref{fig:cLResults}.

\subsection{Investigated samples}
The investigated specimens were cold-rolled steel sheets with a rolling reduction of 70\,\%, resulting in a nominal thickness of approximately 1\,mm. The exact thickness of each specimen was determined using a micrometer and subsequently used in the analysis. Variations between specimens were limited to a few micrometers.
Chemically identical samples (material composition see Table \ref{tab:sample}) were provided in the full-hard (FH) and recrystallized (RX) state. 
From the composition the exact alloy density was estimated, as indicated in Table \ref{tab:sample} and used in the models below.

\begin{figure}
    \centering
    \includegraphics[width=1\linewidth]{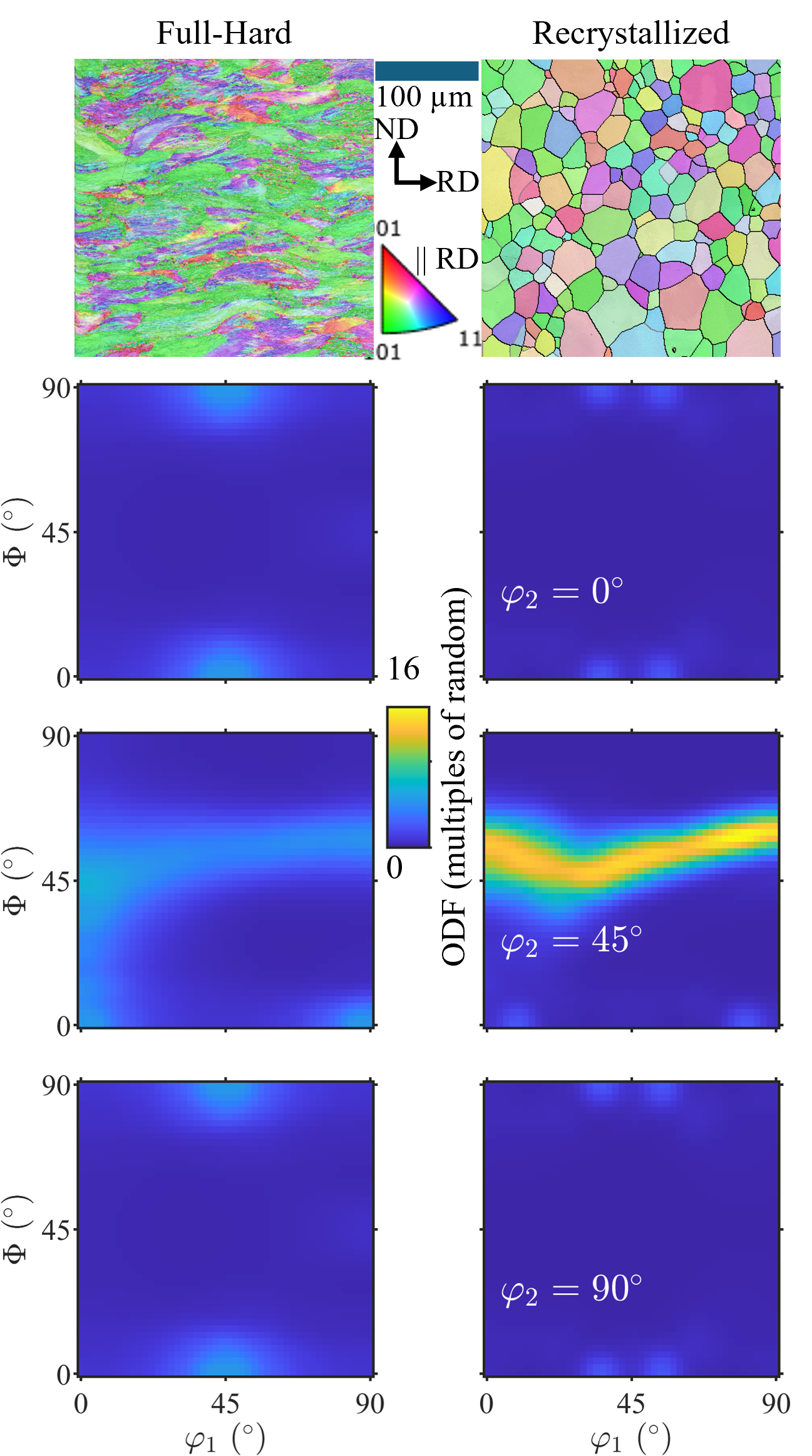}
    \caption{Microstructure of steel samples; Top row: Grain structure reconstructed form EBSD measurements, with grain colors assigned according to inverse pole figure color key, representing the crystallographic directions of the BCC crystal parallel to the rolling direction (RD); Bottom three rows: representative Euler sections of the corresponding ODFs.}
    \label{fig:samplesTexture}
\end{figure}

Representative specimens retained from the same material batches supplied for the laser ultrasound experiments were characterized by electron backscatter diffraction (EBSD) and X-ray diffraction (XRD). The grain structures obtained from EBSD and representative Euler sections derived from the XRD-based orientation distribution functions (ODFs) are shown in Fig. \ref{fig:samplesTexture}. A clear evolution of the grain morphology and crystallographic texture from the FH to the RX condition is evident. As typical for cold-rolled steels, the texture is dominated by the $\gamma$-fibre, with weaker contributions from the $\alpha$-fibre. The RX samples exhibit a more pronounced $\gamma$-fibre texture \cite{Xu2010, Ashby1999}.

For control experiments, a single-crystal silicon wafer with a (100) surface orientation and a thickness of 0.725\,mm was used.

\begin{table}[htbp]
\centering
\caption{Chemical composition of investigated IF-steel grade, and densities used to determine alloy density}
\label{tab:sample}
\begin{tabular}{lcc}
\hline
Element & wt.\,\% & Density (g/cm$^3$) \\
\hline
Fe & 99.6119 & 7.874 \\
Mn & 0.1500 & 7.210 \\
Ti & 0.0800 & 4.510 \\
Al & 0.0470 & 2.700 \\
Cr & 0.0270 & 7.190 \\
Ni & 0.0120 & 8.900 \\
\hline
Alloy (calculated) &  & 7.856 \\
\hline
\end{tabular}
\end{table}

\subsection{Experimental results}
\begin{figure*}
    \centering
    \includegraphics[width=1\linewidth]{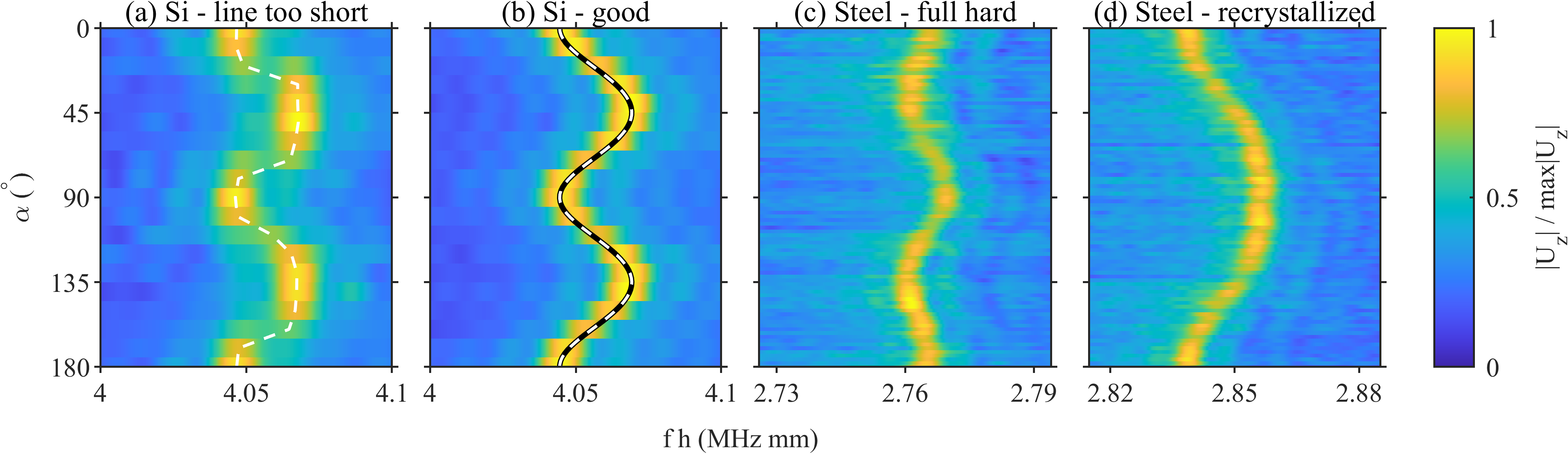}
    \caption{Measured directional ZGV spectra; (a) and (b): Control experiments on Si establishing a suitable line source length; (c) and (d): typical results on steel samples of different texture as indicated.}
    \label{fig:angular_ZGV_scans}
\end{figure*}

The angular scans of the ZGV peaks are shown in Fig.~\ref{fig:angular_ZGV_scans}. To validate the measurement procedure, control experiments were performed on a silicon wafer (0.725 mm thickness, (100) orientation), reproducing the results reported in \cite{Kiefer2023}. These measurements highlight the importance of employing a sufficiently long line source.

Figure~\ref{fig:angular_ZGV_scans}(a) shows results obtained without the 90$^\circ$ cylindrical lens. In this case, the directional selectivity is insufficient, causing the dominant ZGV directions to contribute significantly to neighbouring angles. Consequently, the extracted peak positions (overlaid as white dashed line) form an approximately rectangular curve and exhibit a strong dependence on the selected time window. Under such conditions, misleading angular trends can easily be obtained.

The results shown in Fig.~\ref{fig:angular_ZGV_scans}(b) were achieved using the improved directional selectivity as described. In this configuration, the angular positions of the maxima become independent of the chosen time window, while only the peak amplitudes remain affected, as discussed in \cite{Kiefer2023}. 
The overlaid curve is a calculation of ZGV frequencies for cubic silicon.

All measurements on the steel samples were therefore performed using this approach. It should be noted that the ZGV wavelength in the steel samples (thickness $\sim$\,1 mm) was approximately 1.5 times larger than in the silicon control experiment. Consequently, the minimum required line length is expected to be approximately 1.5 times larger as well. In practice, the line source was chosen considerably longer than this minimum estimate. Furthermore, no indication of insufficient line length was observed, as evidenced by the absence of any time-window dependence of the extracted angular peak positions.
Typical results obtained for the steel samples are shown in Fig.~\ref{fig:angular_ZGV_scans}(c) for the FH state and in Fig.~\ref{fig:angular_ZGV_scans}(d) for the RX state. The angle $\alpha$ denotes the orientation of the generation line with respect to the rolling direction. Consequently, $\alpha=90^\circ$ corresponds to wave propagation parallel to the rolling direction, while $\alpha=0^\circ$ corresponds to propagation transverse to the rolling direction.

\begin{figure}
    \centering
    \includegraphics[width=1\linewidth]{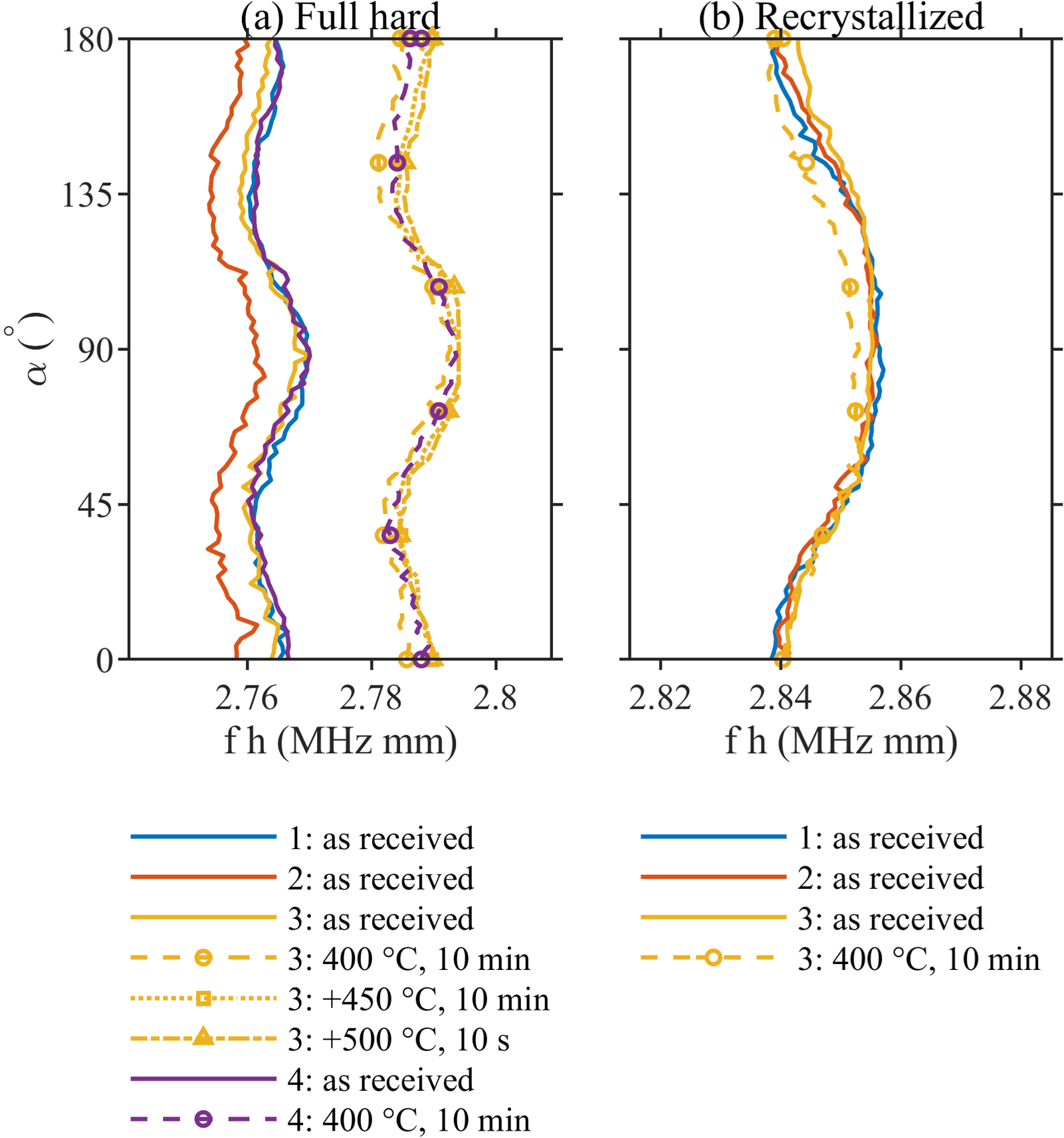}
    \caption{Summarized results for multiple independent samples measured. Full-hard samples in (a) show a  shift in frequency induced by recovery annealing as indicated. Recrystallized samples (b) show excellent repeatability and are unaffected by recovery.}
    \label{fig:repeat}
\end{figure}

Measurements were repeated on independently prepared samples. The extracted ZGV curves, represented by the angular dependence of the peak positions, are summarized in Fig.~\ref{fig:repeat}. For the RX condition, shown in Fig.~\ref{fig:repeat}(b), excellent repeatability was achieved. The FH condition (Fig.~\ref{fig:repeat}(a)) exhibited one sample (sample 2) with a slight deviation from the otherwise consistent results. For each specimen, the thickness was measured using a micrometer screw gauge. Minor inaccuracies in the thickness determination may explain the observed deviation. It should be noted, however, that this relative deviation is only a few permille, indicating very good overall reproducibility.

For the FH condition Fig.~\ref{fig:repeat}(a), additional curves are included for samples that were recovery annealed under the temperatures and holding times indicated in the legend. All recovery treatments resulted in nearly identical ZGV curves despite the different annealing conditions. Compared to the untreated FH state, the curves are shifted to higher frequencies; however, their angular dependence remains essentially unchanged. 
As a control experiment, one RX sample was subjected to the same recovery treatment. In this case, no significant change in the ZGV curve was observed. This suggests that the frequency shift observed for the FH material is associated with a reduction of dislocations present only in the full-hard microstructure. This is corroborated by a quantitative comparison of the observed shift with estimated theoretical values in section \ref{sec:recovery}).

\section{Elastic modelling}
\subsection{Models for elastic stiffness tensor homogenization}

The prediction of macroscopic elastic properties for polycrystals with crystallographic texture requires the single-crystal elastic constants and a statistical representation of the grain orientation distribution. The latter is described by the orientation distribution function (ODF).
The single-crystal elastic constants of body-centered cubic (BCC) iron were taken from \cite{Dever1972}.
ODFs derived from both electron backscatter diffraction (EBSD) and X-ray diffraction (XRD) measurements.  The XRD-derived ODFs yielded better agreement with the experimental results and were therefore used throughout this work. This improved agreement is attributed in part to the finer angular discretization available for the XRD-based ODFs, whereas the ODFs obtained from EBSD were limited by the angular resolution of the available analysis workflow. Additionally, XRD samples a substantially larger grain population, which may further contribute to the observed differences.
\subsubsection{Basic mean field models: Voigt, Reuss and Hill}
For simplicity and reproducibility, standard Voigt, Reuss, and Hill averages were calculated using the freely available MTEX toolbox \cite{MTEX,Bachmann2010}. The Euler angles and corresponding orientation densities obtained from the XRD measurements were used to reconstruct continuous ODFs. Based on these ODFs and the single-crystal elastic constants, the texture-dependent effective stiffness tensors were calculated. 

\subsubsection{Finite element models of representative volume elements: SUBC and KUBC}

The finite element (FE) homogenization approaches considered in this work can be regarded as microstructure-resolved extensions of the classical Voigt and Reuss models. Two complementary FE models were employed: the kinematic uniform boundary condition (KUBC) model and the static uniform boundary condition (SUBC) model. Similar to the Voigt model, KUBC is based on prescribed macroscopic strain states, whereas SUBC is analogous to the stress-controlled Reuss model. However, unlike the mean-field assumptions underlying Voigt and Reuss averaging, both approaches explicitly account for local mechanical interactions within the polycrystalline microstructure. Consequently, the resulting bounds on the effective elastic tensor are expected to be significantly narrower than the corresponding Voigt and Reuss bounds.

\begin{figure}
    \centering
    \includegraphics[width=1\linewidth]{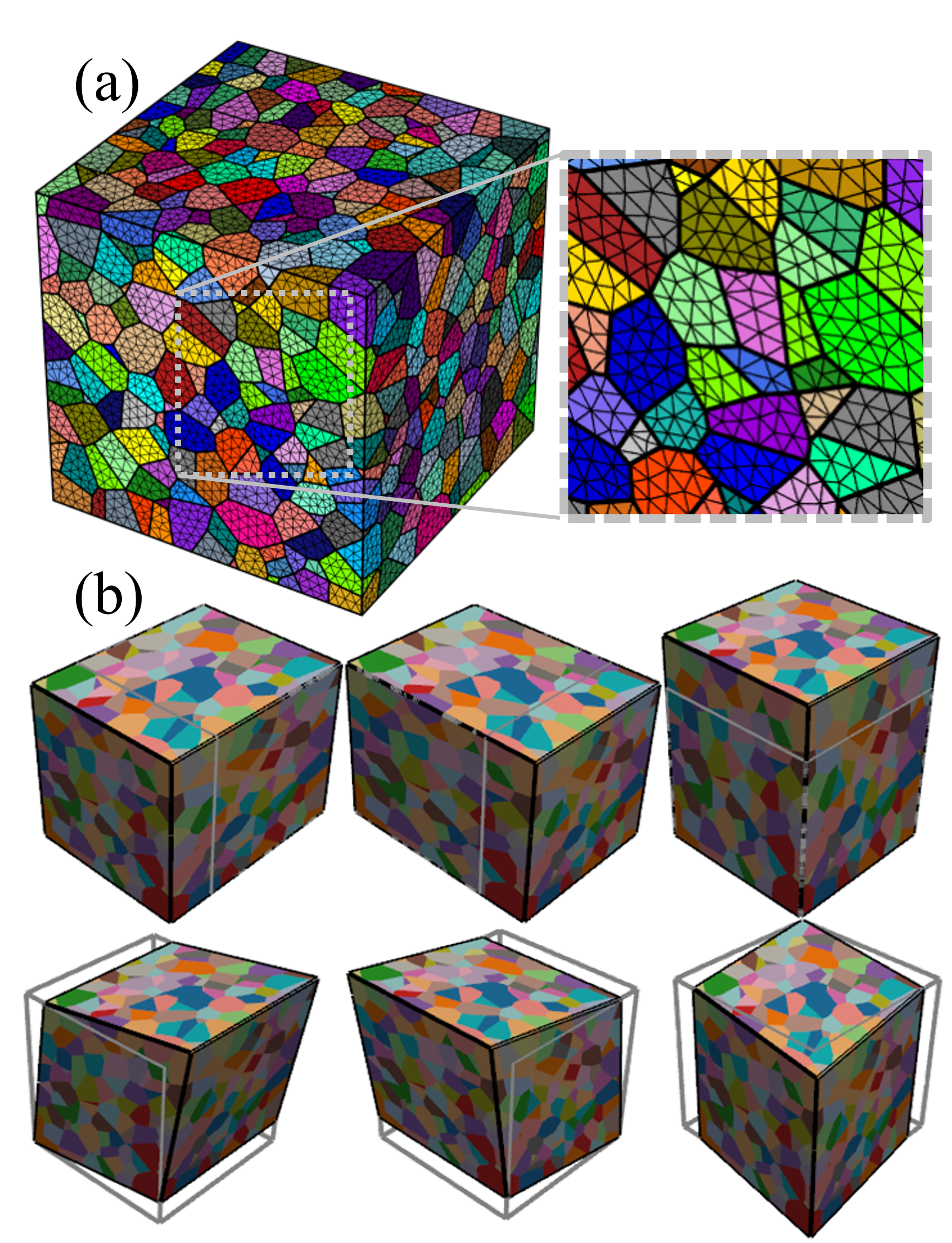}
    \caption{Finite element models; (a) Voronoi cell representation of grains and FE mesh; (b) Illustration of uniform strain states applied to obtain stiffness tensor in KUBC model}
    \label{fig:model}
\end{figure}

The FE homogenization approach is based on a geometrical representation of the polycrystalline microstructure. A synthetic polycrystalline tessellation was generated using Neper \cite{Quey2011,Quey2022} and subsequently discretized by a three-dimensional finite element mesh. Each cell of the tessellation represents an individual grain, such that all finite elements belonging to the same grain share the same crystallographic orientation. An example of the resulting tessellation and finite element mesh is shown in Fig.~\ref{fig:model}(a).

For each grain, an orientation was sampled from the experimentally determined XRD-based ODF and assigned to the corresponding grain of the synthetic microstructure. The single-crystal stiffness tensor of BCC iron was then rotated from the crystal coordinate system into the sample coordinate system according to the assigned Euler angles. 
In this manner, a spatially varying elastic stiffness tensor \begin{equation} \mathbf{C}(\mathbf{x})=\mathbf{C}^{(g)} \end{equation} was obtained, where $\mathbf{C}^{(g)}$ denotes the rotated single-crystal stiffness tensor of grain $g$. The polycrystal was modeled as a heterogeneous linear elastic material with a spatially varying stiffness tensor. The finite element simulations were implemented in the open-source FEniCSx framework \cite{Baratta2023}, and the displacement field was discretized using quadratic finite elements.

To determine the complete effective elastic tensor of the polycrystal, six independent macroscopic loading states were applied. Figure~\ref{fig:model}(b) illustrates the corresponding homogeneous deformation modes using distorted cubes. Since each applied unit loading state produces all six components of the macroscopic response, one simulation determines one column of the effective stiffness or compliance matrix. Consequently, six simulations are sufficient to determine the full $6\times6$ elastic tensor. 
For the KUBC approach, homogeneous strain states were imposed on the external boundary of the representative volume element. The resulting volume-averaged stresses directly provide the columns of the effective stiffness tensor. In contrast, the SUBC approach applies homogeneous stress states through boundary tractions. The resulting volume-averaged strains yield the columns of the effective compliance tensor, which is subsequently inverted to obtain the effective stiffness tensor.

Additional details on the model are found in the appendix \ref{app:fe}. An example implementation is  available within the data archive complementing this article.

\begin{figure}
    \centering
    \includegraphics[width=1\linewidth]{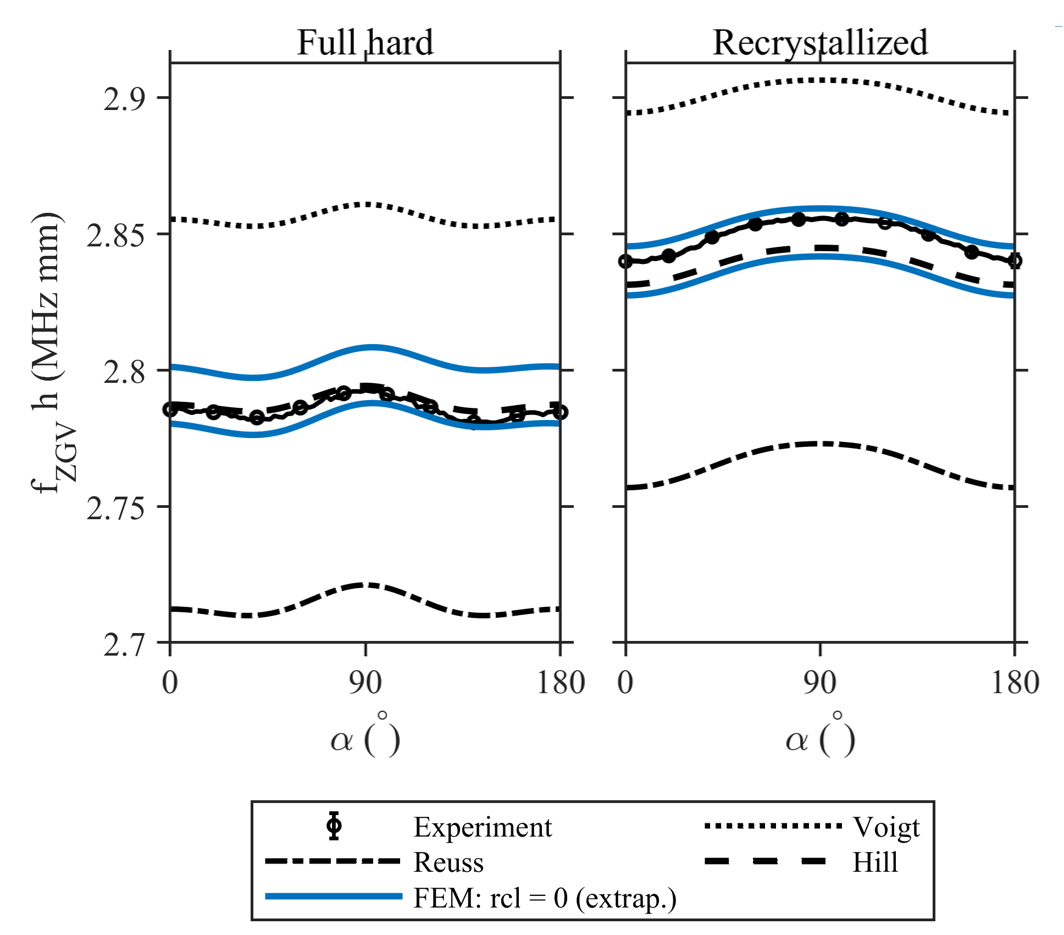}
    \caption{Comparison of representative models and experiments, highlighting the broad Voigt-Reuss range vs. the narrow SUBC-KUBC (FEM) range}
    \label{fig:modelsOverview}
\end{figure}

Figure~\ref{fig:modelsOverview} provides an overview of the final ZGV frequency predictions for both investigated material states. The Voigt, Reuss, and Hill estimates as well as the FE results all reproduce the characteristic angular dependence of the measured ZGV frequencies. The main difference lies in the width of the predicted bounds: the classical Voigt and Reuss estimates form a comparatively wide interval, whereas the corresponding KUBC and SUBC results are much more closely spaced. For both material states, the Hill prediction and the experimental results lie within the respective bounds. The FE curves shown in Fig.~\ref{fig:modelsOverview} correspond to values extrapolated towards an infinitely fine finite element discretization. 

\begin{figure}
    \centering
    \includegraphics[width=1\linewidth]{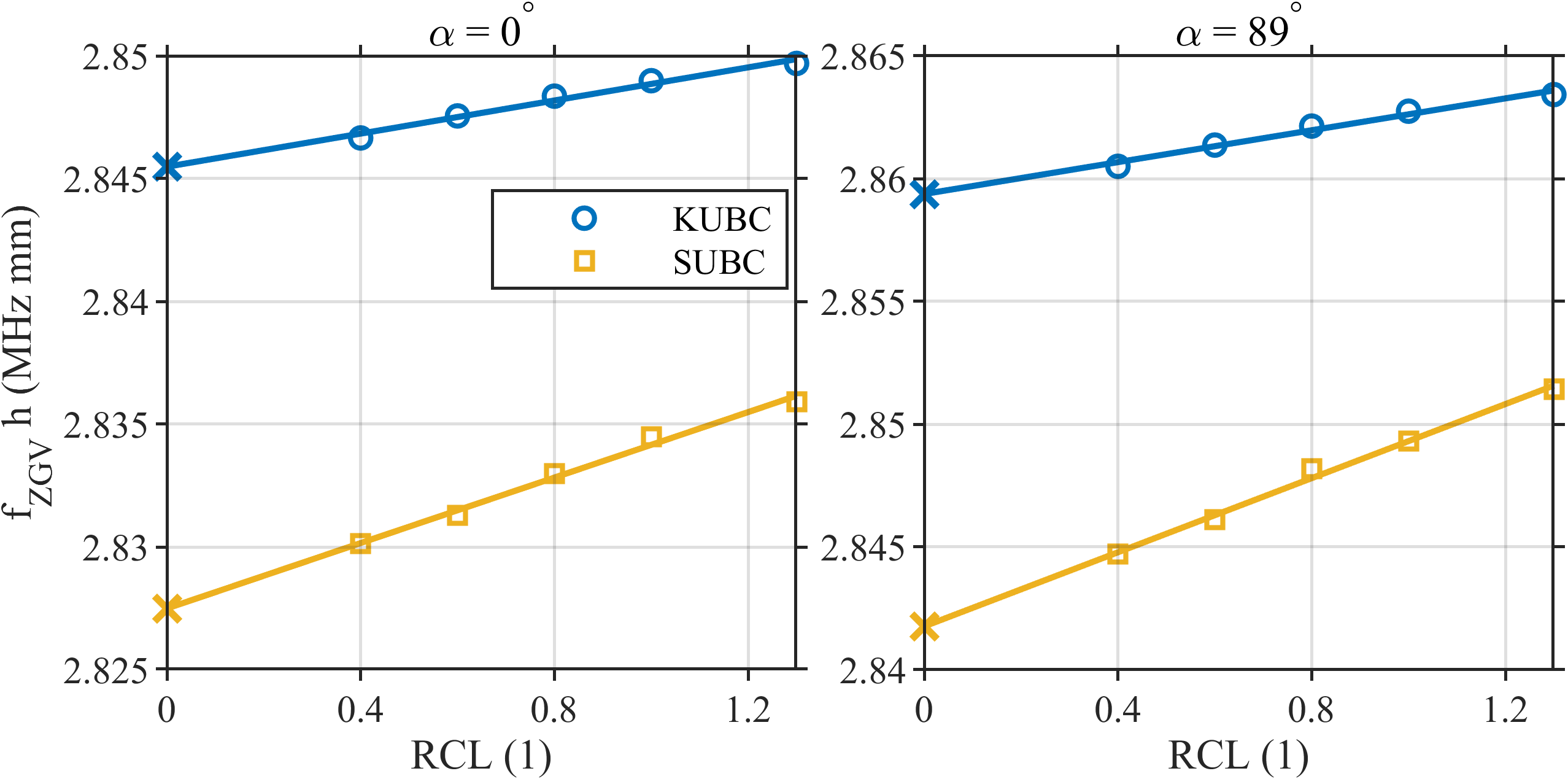}
    \caption{FEM model mesh convergence and extrapolation to relative characteristic lengths (RCL) of 0, shown for RX texture, 2000 grains}
    \label{fig:extraprcl0}
\end{figure}

\begin{figure}
    \centering
    \includegraphics[width=1\linewidth]{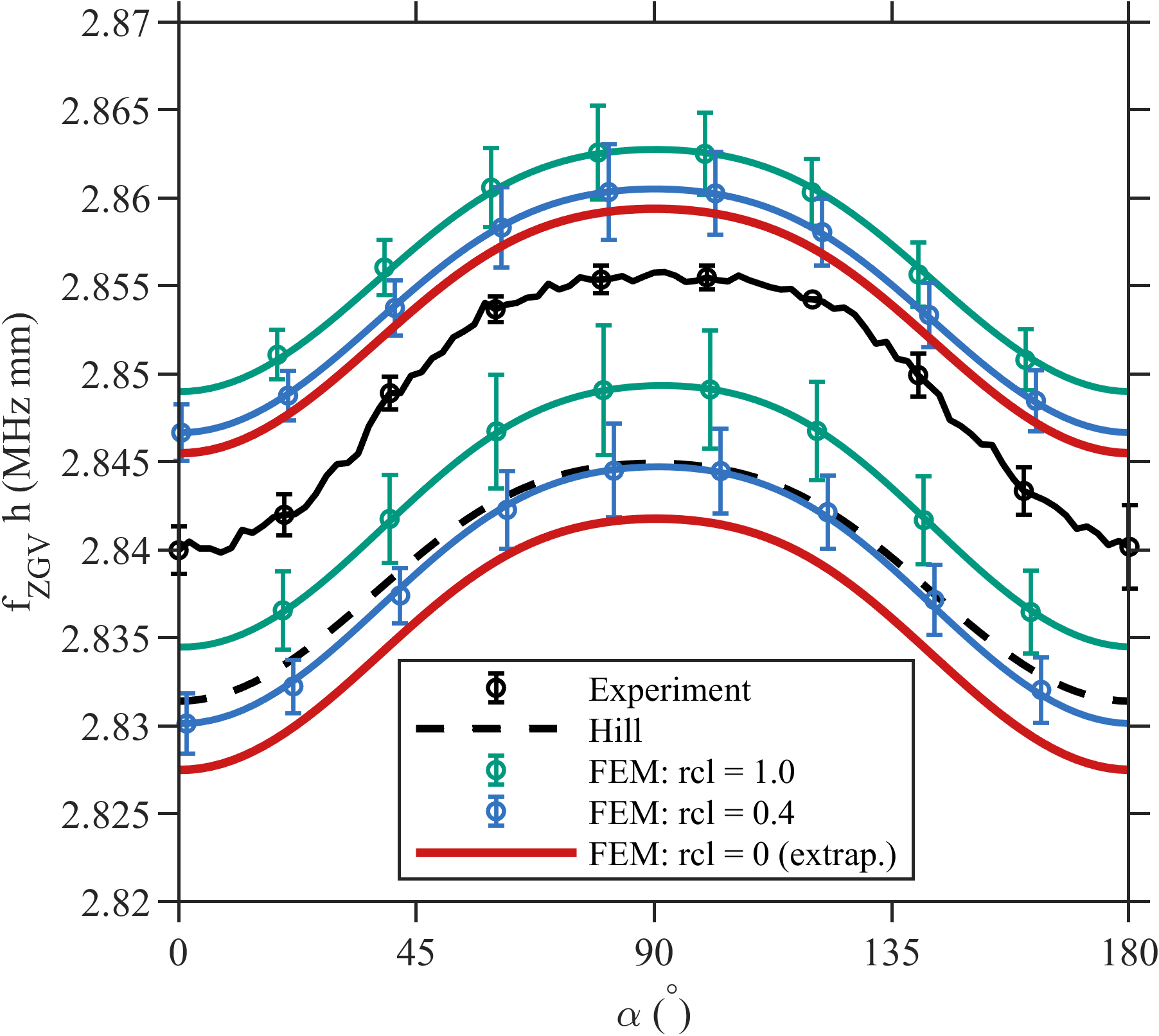}
    \caption{Modeled and measured ZGV frequencies for the RX case, zoom on SUBC-KUBC range and influence of mesh finesse; experiments and Hill-average fall within the bounds of the RCL=0 FEM models (red curves)}
    \label{fig:KGZoom}
\end{figure}

To assess the influence of the mesh resolution, the simulations were repeated for several values of the relative characteristic length (RCL) used by Neper to control the target element size. In three dimensions, the default value $\mathrm{RCL}=1$ corresponds to approximately 100 finite elements per grain, while smaller RCL values result in progressively finer meshes. The simulations considered RCL values between 1.3 and 0.4. A small systematic dependence of the predicted ZGV frequencies on RCL remained.To provide an estimate of a mesh independent result ($\mathrm{RCL}=0$), the ZGV frequencies were extrapolated directly as a function of RCL. For each propagation angle, a linear fit was applied independently to the results obtained at the different mesh resolutions, and the value at $\mathrm{RCL}=0$ was taken as the extrapolated result corresponding to an infinitely fine discretization. 
This may be regarded as a Richardson-type extrapolation \cite{Roy2005}, assuming a first-order dependence on the characteristic mesh size. A more extensive numerical convergence study was beyond the scope of this work. While the extrapolated values do not constitute a rigorous proof of convergence, the good agreement with the experimental results supports the extrapolated predictions.
The procedure was performed separately for the KUBC and SUBC results. Figure~\ref{fig:extraprcl0} illustrates this extrapolation for $\alpha=0^\circ$ and $\alpha=90^\circ$ for the RX material. 
Figure~\ref{fig:KGZoom} provides a more detailed view of the FE results in comparison with the experimental data and the Hill prediction. To illustrate the mesh-resolution dependence, the results for $\mathrm{RCL}=1.0$ and $\mathrm{RCL}=0.4$ are shown together with the extrapolated $\mathrm{RCL}=0$ curves. The error bars represent the variation obtained by repeating the simulations with different sets of grain orientations sampled from the experimentally determined ODF.
For the experimental data in Figure~\ref{fig:KGZoom}, the error bars come from the variation between the different experimental KG samples.

\begin{figure}
    \centering
    \includegraphics[width=1\linewidth]{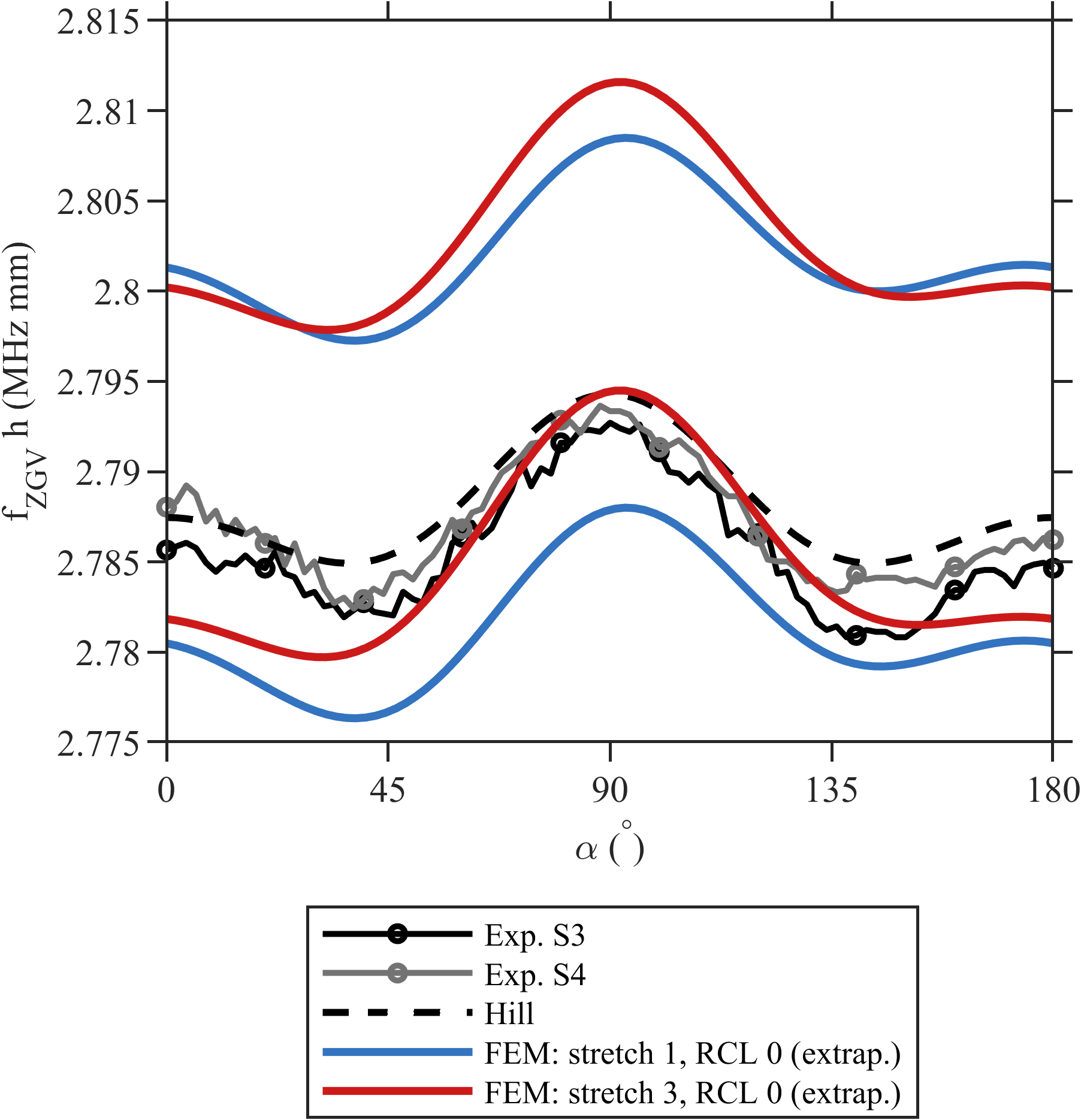}
    \caption{Modeled and measured (sample 3 and sample 4 in the recovered state) ZGV frequencies for the FH case, zoom on SUBC-KUBC range and influence of grain morphology; experiments and Hill-average fall within the bounds of the RCL=0 FEM models with equiaxial grains (stretch=1, blue curves)}
    \label{fig:WHZoom}
\end{figure}

For the FH material, Figure~\ref{fig:WHZoom} provides a closer view of the extrapolated FE results in comparison with the experimental data and the Hill estimate. To account for the elongated grain morphology introduced by rolling, an additional stretched representative volume element was considered. The grain shape was modified directly in Neper by imposing an aspect ratio corresponding approximately to the macroscopic deformation associated with a rolling degree of 70\%. This was represented by a compression by a factor of 3 in the thickness direction ($z$) and a corresponding elongation by a factor of 3 in the rolling direction ($x$). In addition to the equiaxed reference case, the extrapolated KUBC and SUBC predictions for this stretched microstructure are therefore included in Figure~\ref{fig:WHZoom}, providing a slightly different result.

\begin{figure}
    \centering
    \includegraphics[width=1\linewidth]{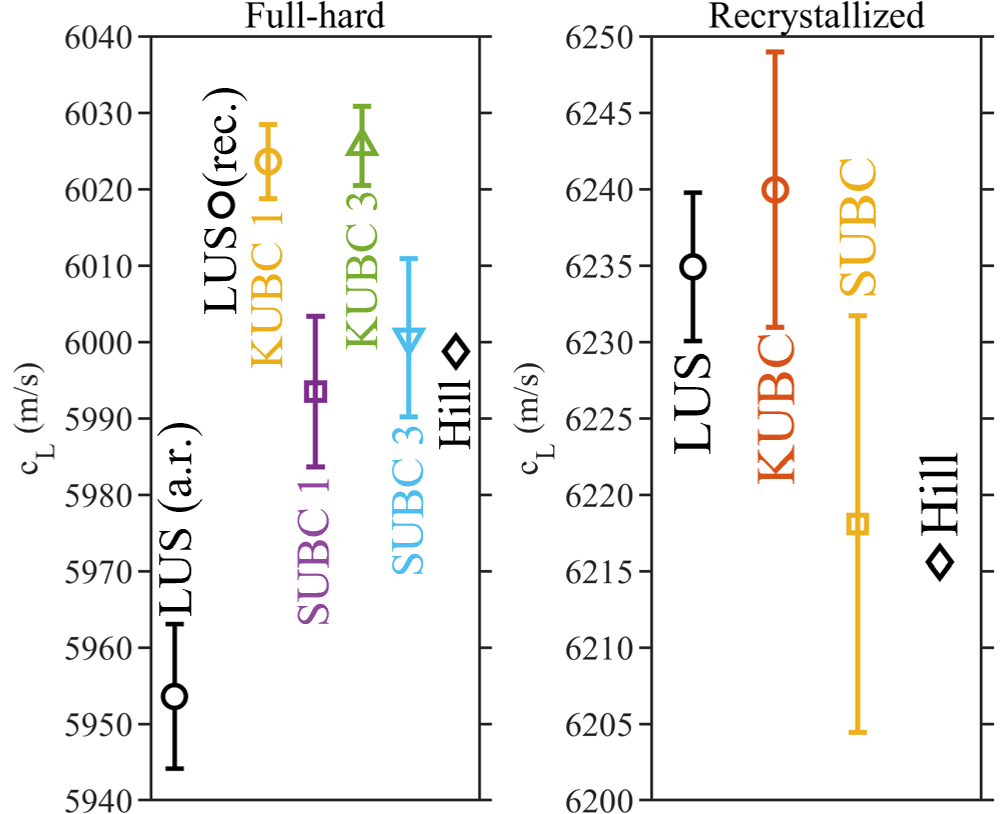}
    \caption{Measured and simulated values for $c_\mathrm{L}$ in normal direction (though-thickness); for the FH case (left panel) LUS results on samples as received (LUS a.r.) and recovery annealed samples (LUS rec.) are included, as well as FE results with equiaxed (KUBC 1 / SUBC 1) and 3 fold elongated/compressed microstructure (KUBC 3 / SUBC 3).}
    \label{fig:cLResults}
\end{figure}

From the modeled stiffness tensors we also calculated the longitudinal sound velocity $c_\mathrm{L}$ in through-thickness direction. The results are shown together with experimental values in Fig. \ref{fig:cLResults}

\section{Discussion}
\subsection{Comparison of experiment and models}

Careful experimental procedures, including sufficiently long line sources as discussed in Fig.~\ref{fig:angular_ZGV_scans}, removal of surface oxides, and local thickness measurements, resulted in measurements that are in close agreement with the predictions based on material homogenization. This agreement is observed for both the ZGV frequencies and the longitudinal wave velocity $c_\mathrm{L}$ in the through-thickness direction.

For the material model, the elastic constants of pure cubic BCC iron were used. The total alloying content of the investigated material was below approximately 0.4\%, and its influence was therefore neglected in the elastic stiffness tensor. The actual alloy composition was, however, taken into account in the material density.

The FE homogenization based on representative volumes containing 2000 grains yields comparatively narrow KUBC/SUBC bounds. For the ZGV frequency, the separation between the two FE bounds is approximately 0.7\% of the absolute frequency, compared with approximately 5\% between the classical Voigt and Reuss bounds. For both investigated material states, the experimental results lie within the corresponding FE bounds.

For the RX material, which showed excellent experimental repeatability and an approximately equiaxed grain morphology, the measured response is about 0.3\% above the Hill prediction. Overall, the Hill average nevertheless provides a remarkably close approximation to the experimental response.

Interestingly, the Hill average also provides a good approximation for the strongly textured and elongated FH material. This is not necessarily expected a priori, since the Hill estimate does not explicitly account for the spatial arrangement and morphology of the grains. The experimental results obtained after recovery annealing below the RX temperature also remain within the predicted FE bounds.

\subsection{Influence of grain morphology}

The FE model additionally allows the influence of grain elongation to be investigated. Comparing the equiaxed morphology (stretch~1) with the elongated morphology (stretch~3) reveals only a relatively small effect, particularly for the SUBC results; see Fig.~\ref{fig:WHZoom}. The angular variation observed experimentally is in fact somewhat closer to the prediction obtained for the unstretched morphology. The influence of grain elongation may therefore be smaller than the width of the present numerical bounds.

We note that, in the elongated model, each grain is discretized by substantially more FE elements along the rolling direction than along the transverse/normal direction. It has proven difficult to exclude the possibility that this introduces a numerical anisotropy.

Nevertheless, an influence of grain morphology on the effective elastic response appears intuitive when considering the idealized pictures underlying the Voigt and Reuss limits. Along the rolling direction, strongly elongated grains resemble a more columnar arrangement, in which many grains experience a similar deformation state, qualitatively approaching the Voigt picture. In the compressed direction, the grains are arranged more nearly in series, which is closer to the Reuss picture \cite[Figure~3]{Bunge2000}. 

Further investigation of the relation between grain aspect ratio and the effective elastic response would therefore be of both scientific interest and technological relevance. In particular, grain-shape sensitivity could provide an additional ultrasonic indicator of recrystallization in cases where the crystallographic texture changes only weakly \cite{Diehl2019, Militzer2013, Ren2025}.

\subsection{Effect of recovery annealing for FH samples}
\label{sec:recovery}
For the FH material, a systematic shift in the ultrasonic response was observed after recovery annealing, as shown in Fig.~\ref{fig:repeat} and Fig.~\ref{fig:cLResults}. The experimental results for the annealed material remains within the FE bounds. This suggests that the as-received material differs from the recovered material in a way that is not captured by the FE model. 

Our hypothesis is, that this is due to dislocations introduced by cold-rolling which are reduced by recrystallization (note that recovery treatment had no measurable effect on the RX samples) are reduced by three orders of magnitude, from $\sim 10^{15}$ to $\sim 10^{12}\,\mathrm{m}^{-2}$. We estimate the recovery annealing below the RX temperature is known to remove approximately 30\% of dislocations, while having no effect on the crystallographic texture \cite{Sugiyama2021, Smith2004, Williamson1956}. 

Following the low-frequency approximation of Maurel et al.\ \cite{Maurel2005}, the dislocation-induced velocity changes $\Delta c_\mathrm{L}$ and $\Delta c_\mathrm{T}$ are estimated as

\begin{align}
    \label{eq:GLapprox_cT}
    \frac{\Delta c_\mathrm{T}}{c_\mathrm{T,rec}}
    &\simeq
    \frac{8}{5\pi^3}\Lambda L^2,    \\
    \label{eq:GLapprox_cL}
    \frac{\Delta c_\mathrm{L}}{c_\mathrm{L,rec}}
    &\simeq
    \frac{32}{15\pi^3}
    \left(\frac{c_\mathrm{T}}{c_\mathrm{L}}\right)^2
    \Lambda L^2.
\end{align}

Here, \(\Lambda\) denotes the \textit{movable} dislocation density, i.e.\ the total length of \textit{acoustically movable} dislocations per unit volume, and \(L\) the effective free length of the pinned dislocation segments. 

Note, that the literature-based values listed above refer to \textit{total} dislocation densities. Since our measured quantities after recovery agree well with the model predictions, which do not account for a dislocation contribution, we assume that the density of acoustically movable dislocations is reduced to approximately zero, \(\Lambda_{\mathrm{rec}}\approx 0\). We further assume that the dislocations removed during recovery correspond to the acoustically movable fraction of the total dislocation population. This gives \(\Lambda_{\mathrm{ar}}\approx 3\times10^{14}\,\mathrm{m^{-2}}\).

From the measured \(c_\mathrm{L}\) and the mean of \(f_{\mathrm{ZGV}}(\alpha)\), we estimate isotropic sound velocities for the recovered and as-received states, resulting in \(c_\mathrm{L,rec}=6017.9\,\mathrm{m\,s^{-1}}\), \(c_\mathrm{T,rec}=3235.0\,\mathrm{m\,s^{-1}}\), and \(\Delta c_\mathrm{L}=55.6\,\mathrm{m\,s^{-1}}\) and \(\Delta c_\mathrm{T}=60.6\,\mathrm{m\,s^{-1}}\).

Insertion into Eqs.~(\ref{eq:GLapprox_cT}) and (\ref{eq:GLapprox_cL}) yields effective free dislocation lengths of \(L_\mathrm{T}\approx34.8\,\mathrm{nm}\) and \(L_\mathrm{L}\approx39.4\,\mathrm{nm}\). These independently obtained values are in good agreement, and the resulting absolute length scale appears realistic for the dense dislocation structure of the 70\% cold-rolled steel. The observed shifts in sound velocity are thus likely related to a change in dislocation density.

\subsection{Parameter identifiability}

While the information content of directional ZGV measurements (Fig.\,\ref{fig:angular_ZGV_scans}) appears to be high, the extent to which the elastic properties of a specimen can actually be constrained from such measurements is not obvious. 
We therefore analyze the identifiability of parameters  based on a local perturbation of the forward model (see Appendix \ref{app:sensitivity_theory}) and \cite{Tarantola2005inverse}. 
In this approach, changes in the observables $\mathbf{y}$ (i.e. ZGV frequencies in different directions) are linearly related to changes in the parameters $\mathbf{p}$ (i.e. the independent entries of the elastic tensor) by the (normalized) sensitivity matrix $\mathbf{J}$, which can be approximated by simple numerical differentiation. 

A singular value decomposition of $\mathbf{J}$ then orders independent (orthonormal) combinations of input parameters changes by their influence on the observables, given by their corresponding singular value $\sigma_i$. Singular values below a chosen threshold are considered $0$, and the corresponding combinations of parameter changes leave the observables unchanged. The number of non-zero singular values yields the effective rank of $\mathbf{J}$. 
Only if the the rank matches the number of parameters, all parameters can uniquely be found in an inverse problem. Otherwise certain combinations remain undetermined (i.e. the parameter-Nullspace).

By normalizing $\mathbf{J}$ (see Appendix \ref{app:sensitivity_theory}), the analysis can be extended to include additional parameters (such as thickness or density) and also additional observables. 
We summarize the obtained ranks for different materials and combinations of LUS measurements in  table \ref{tab:jacobian_ranks}. 
Besides directional ZGV frequencies $f_\mathrm{ZGV}(\alpha)$, we also include through-thickness longitudinal sound velocity $c_\mathrm{L}$ and $c_\mathrm{SAW}(\alpha)$, i.\,e.\ the phase velocity of the lowest order Lamb-modes at a wavenumber where it approximates a surface wave \cite{Stoklasova2021TGS, Watzl2025SAW}. 
These modalities are chosen, as we consider them practical for in-situ investigations. 
They can be achieved with single-sided access to a single specimen, and they could potentially be combined in a single LUS measurement \cite{Watzl2025CoMLAS}. 
In addition to the steel samples investigated here where the reference point is the Hill average, crystalline silicon and an isotropic steel are included.

For each case listed in Table \ref{tab:jacobian_ranks}, two effective ranks are reported, corresponding to different thresholds applied to the singular values. A threshold of $10^{-6}$ is used to assess local theoretical identifiability by excluding numerically negligible singular values, while a threshold of $2 \cdot 10^{-2}$ is adopted to assess practical identifiability, accounting for the limited precision of experimental measurements. For each effective rank, the corresponding condition number $\kappa = \sigma_{\max}/\sigma_{\min}$ is provided in parentheses, where $\sigma_{\min}$ denotes the smallest singular value retained above the respective threshold.

Directional quantities were evaluated over the range from $0^\circ$ to $180^\circ$ in increments of $10^\circ$.

\newcommand{\kappastyle}[1]{{\small #1}} %
\renewcommand{\rankval}[4]{#1 (\kappastyle{#2}) / #3 (\kappastyle{#4})} %

\begin{table*}[htbp]
\centering
\caption{Results of identifiability analysis for different materials and different combinations of experimental modalities. Resulting ranks of the sensitivity matrix are shown for numerical and practical singular value thresholds, with the smallest singular value indicated in brackets.}
\label{tab:jacobian_ranks}
\renewcommand{\arraystretch}{1.2} %
\begin{tabular}{@{} l c c c c @{}}
\toprule
\multirow{2}{*}{\textbf{Material}} & \textbf{Independent} & \multicolumn{3}{c}{\textbf{Sensitivity Matrix Rank} ($\sigma \ge 10^{-6}$ / $\sigma \ge 2\cdot 10^{-2}$)} \\
\cmidrule(l){3-5}
& \textbf{Tensor Entries} & ZGV($\alpha$) & ZGV($\alpha$) + $c_\mathrm{L}$ & ZGV($\alpha$) + $c_\mathrm{L}$ + $c_\mathrm{SAW}$($\alpha$) \\
\midrule

Steel FH & 9 
& \rankval{6}{$2.1\times 10^5$}{3}{35.7} 
& \rankval{7}{$2.4\times 10^5$}{4}{37.7} 
& \rankval{9}{$4.9\times 10^3$}{7}{50.4} \\

Steel RX & 9 
& \rankval{6}{$4.4\times 10^5$}{3}{30.0} 
& \rankval{7}{$5.8\times 10^5$}{4}{31.8} 
& \rankval{9}{$7.2\times 10^5$}{6}{40.9} \\

Silicon & 3 
& \rankval{3}{$8.6\times 10^3$}{2}{91.1} 
& \rankval{3}{$2.4\times 10^2$}{2}{44.9} 
& \rankval{3}{5.5}{3}{5.5} \\

Isotropic Steel & 2 
& \rankval{1}{1.0}{1}{1.0} 
& \rankval{2}{6.3}{2}{6.3} 
& \rankval{2}{1.9}{2}{1.9} \\

\bottomrule
\end{tabular}
\end{table*}

Assuming that thickness and density are known, the identifiability analysis leads to the following conclusions:
\begin{itemize}
    \item Isotropic steel: As expected, all elastic parameters can be identified by combining at least two measurement modalities, consistent with \cite{Clorennec2007_ZGVSoundVelocity}.

    \item Cubic silicon: The elastic parameters can be fully identified under ideal experimental conditions using ZGV measurements alone or in combination with $c_\mathrm{L}$ (as measured in \cite{Prada2009}). However, under the practical identifiability criterion adopted here, full rank is achieved only when $c_\mathrm{SAW}$ measurements are included.

    \item Orthorhombic steels: Full local identifiability is achieved numerically only when all three measurement modalities are combined. Under the practical identifiability criterion, the effective rank is limited to 3 for the investigated ZGV modes and increases to 4 when combined with $c_\mathrm{L}$ measurements. Nevertheless, this demonstrates that ZGV measurements substantially increase the number of independently accessible elastic parameter combinations compared with measurements of $c_\mathrm{L}$ alone.
\end{itemize}

\section{Conclusion}

Directional zero-group-velocity measurements were demonstrated as a repeatable and sensitive means giving insight on the effective elastic anisotropy of textured steel plates. 
The experiments were predicted well by models based on the measured crystallographic texture and homogenized elastic properties. 
Both classical mean-field approaches and microstructure-resolved finite-element homogenization captured the main experimental trends, with the KUBC/SUBC bounds being substantially narrower than the classical Voigt--Reuss bounds. Despite its simplicity, the Hill estimate provided a remarkably good approximation for both the RX and FH material.
A small influence of grain elongation on the effective elastic response was indicated by the FE model. However, the predicted effect was too small to be validated experimentally. 
Further investigation is required to rule out possible numerical artefacts and to test the morphology effect experimentally.

The excellent agreement between experiment and model for the RX samples, contrasted by the systematic mismatch for the FH state (as received), motivated us to investigate a possible dislocation contribution. 
For the FH samples recovery annealing produced a clear shift of the ultrasonic response towards the model prediction, while leaving the angular dependence essentially unchanged. 
The magnitude of this shift is consistent with the order of magnitude expected from a simplified Granato--L\"ucke-type estimate \cite{Maurel2005}, supporting dislocations as the origin of the discrepancy in the FH state.

The presented LUS measurements were assessed regarding parameter identifiability.
Considering practical feasibility, 3 combinations of the 9 parameters are determined from directional ZGV measurements only.
Consequently, certain texture changes may produce a measurable change in the directional ZGV response \(f_{\mathrm{ZGV}}(\alpha)\) while having no effect on through-thickness sound velocity. 
The practical identifiability was increased up to 6 or 7 parameter combinations respectively if through-thickness longitudinal wave velocity \(c_\mathrm{L}\) and the directional surface acoustic wave velocity \(c_\mathrm{SAW}(\alpha)\) were included in addition to the ZGV data. Such a complementary measurement set could be implemented elegantly by extending the previously introduced Combined Mode Local Acoustic Spectroscopy \cite{Watzl2025CoMLAS}. 

For robust recrystallization-monitoring, the directional ZGV response may be normalized by the through-thickness longitudinal velocity, \(f_{\mathrm{ZGV}}(\alpha)/c_\mathrm{L}\), to obtain a thickness-independent angular trend that is characteristic of the elastic anisotropy. An in-situ implementation would be feasible by rotating the cylindrical excitation lens, rather than the sample.

Overall, the results show that directional ZGV measurements provide rich information on the macroscopic elastic anisotropy of textured steel sheets and are well suited for comparison with texture-based elastic models.  A combination with complementary measurements of other mode velocities allows to extend the information content or method robustness. This offers a promising route towards the non-contact determination of anisotropic elastic properties and the monitoring of microstructural changes during thermal processing.

\acknowtext

This work was funded by the Austrian Research Promotion Agency (FFG) within the Bridge programme under the research project \enquote{\foreignlanguage{naustrian}{Laser-Ultraschall-Methoden zum Beobachten metallurgischer Prozesse in Stahl während thermischer Behandlung}} (STAHLUS; Grant No.\ FO999901377), by the Government of Upper Austria within the FTI programme under the project \enquote{\foreignlanguage{naustrian}{Effizienz und QUAlität --- Innovative photonische Methoden zur Materialcharakterisierung und Prozessoptimierung}} (EQUAMP), and by the Austrian Science Fund
(FWF) via the research project \enquote{Guided waves in piezoelectric plates and layers} (GUWAPI; Grant No.\ PAT1973825).
We thank Prof. Matthias Millitzer (University of British Columbia) for valueable discussions throughout the project, and Dr. Daniel Kiefer (Institut Langevin, ESPCI Paris) for providing GEWTool to the research community. 

\dataavailability
The data necessary for reproducing the results in this article are freely available in Zenodo: \href{https://doi.org/10.5281/zenodo.22894167}{https://doi.org/10.5281/zenodo.22894167} \cite{Gruensteidl2026Zenodo}.

\authorcontrib

\textbf{C.~Gr\"{u}nsteidl}: Conceptualization (lead); Writing – original draft (lead); Writing – Review \& Editing (equal); Project administration (lead); Supervision (equal); Funding acquisition (equal); Investigation (lead);  Software (lead); Visualization (lead); Methodology (lead). %
\textbf{C.~Kerschbaummayr}:  Investigation (equal); 
\textbf{G.~Yan}: Writing – Review \& Editing (supporting); Investigation (supporting); Validation (supporting). %
\textbf{G.~Watzl}: Writing – Review \& Editing (supporting); Data Curation (lead); Software (supporting); Investigation (supporting). %
\textbf{F.~Rudinger}: Writing – Review \& Editing (supporting); Investigation (supporting). %
\textbf{E.~Scherleitner}: Writing – Review \& Editing (supporting); Resources (equal). %
\textbf{M.~Sonnleitner}: Conceptualization (equal); Writing – Review \& Editing (supporting); Resources (equal).
\textbf{C.~Hoflehner}: Conceptualization (equal); Writing – Review \& Editing (supporting); Investigation (equal); Resources (equal). %

\appendix
\section{Finite element implementation details}
\label{app:fe}

The polycrystalline representative volume elements were generated using Neper \cite{Quey2011,Quey2022} and imported into the open-source FEniCSx finite element framework \cite{Baratta2023}. Grain identifiers associated with the tessellation were used to assign crystallographic orientations and material properties on an element-wise basis.

The elastic behaviour was modeled using infinitesimal linear elasticity. For each grain, the single-crystal stiffness tensor of BCC iron was rotated from the crystal coordinate system into the sample coordinate system according to the assigned Euler angles. The resulting stiffness tensor was assumed constant within each grain and was assembled as a piecewise constant spatial field over the computational domain.

The displacement field was discretized using quadratic Lagrange finite elements. The weak form of the equilibrium equations was implemented using the Unified Form Language (UFL) \cite{Alnaes2014} and assembled with DOLFINx. Parallel execution was achieved using MPI.

For the KUBC simulations, the displacement field on the outer boundary of the RVE was prescribed according to

\begin{equation}
\mathbf{u}(\mathbf{x})
=
\mathbf{E}^{M}\mathbf{x},
\end{equation}

where $\mathbf{E}^{M}$ denotes the prescribed macroscopic strain tensor. Six independent loading cases corresponding to the normal and shear strain components in Voigt notation were analyzed. The effective stiffness tensor was obtained from the resulting volume-averaged stresses,

\begin{equation}
\langle \boldsymbol{\sigma} \rangle
=
\frac{1}{V}
\int_V
\boldsymbol{\sigma}
\,\mathrm{d}V.
\end{equation}

For the SUBC simulations, a homogeneous traction field

\begin{equation}
\mathbf{t}
=
\boldsymbol{\Sigma}^{M}\mathbf{n}
\end{equation}

was applied to the external boundary, where $\boldsymbol{\Sigma}^{M}$ is the prescribed macroscopic stress tensor and $\mathbf{n}$ is the outward unit normal vector. To eliminate rigid-body motion associated with the resulting Neumann problem, additional displacement constraints were imposed on selected corner nodes of the RVE. Six independent stress loading cases were solved and the resulting volume-averaged strains,

\begin{equation}
\langle \boldsymbol{\varepsilon} \rangle
=
\frac{1}{V}
\int_V
\boldsymbol{\varepsilon}
\,\mathrm{d}V,
\end{equation}

were assembled into the effective compliance tensor. The corresponding effective stiffness tensor was subsequently obtained by matrix inversion.

The resulting linear systems were solved using iterative PETSc solvers with algebraic multigrid preconditioning. All calculations were performed in parallel using distributed-memory domain decomposition.

\section{Appendix B: Implementation of the Identifiability analysis}
\label{app:sensitivity_theory}

Let \(\mathbf p\) denote the vector of model parameters, which in the present case consists of the \(n\) independent entries of the macroscopic stiffness tensor, and let \(\mathbf y\) denote the vector of \(m\) measured quantities. 
In the present study, the entries of \(\mathbf y\) correspond primarily to ZGV frequencies measured for different propagation directions (and, where applicable, different ZGV modes). 
Additional observables can readily be included by augmenting the measurement vector. 
Linearization of the forward model about a reference parameter set yields
\begin{equation}
    \Delta\mathbf y = \mathbf{J}\Delta\mathbf p,
\end{equation} 
where 
\begin{equation}
    \mathbf{J} = \left[
        \frac{\partial\mathbf y}{\partial p_1} 
        \ \dots\  
        \frac{\partial\mathbf y}{\partial p_n} 
    \right]
\end{equation}
is the \(m\times n\) Jacobian matrix. 
The Jacobian is evaluated numerically by perturbing the model parameters and computing the corresponding changes in the predicted measurements. 
To remove the influence of parameter and measurement scaling, the analysis is performed using normalized sensitivities. 
The resulting dimensionless formulation enables observables with different units and magnitudes to be combined within a single sensitivity matrix. 
The elements of the normalized Jacobian are defined as 
\begin{equation}
    \tilde J_{ij} = \frac{p_j}{y_i} \frac{\partial y_i}{\partial p_j},
\end{equation} such that \(\tilde J_{ij}\) represents the relative change in observable \(y_i\) resulting from a relative change in parameter \(p_j\). 
This formulation is independent of the specific choice of measurements and therefore permits the direct inclusion of additional observables, such as through-thickness longitudinal wave velocities \(c_\mathrm{L}\), to evaluate their contribution to parameter identifiability. 
The information content of the measurement set is then assessed through the singular value decomposition (SVD) 
\begin{equation}
    \tilde{\mathbf{J}} = \mathbf{U}\mathbf{\Sigma}\mathbf{V}^\dagger\, 
\end{equation}    
where \(\mathbf{U}\) is an \(m\times m\) orthogonal matrix containing the left-singular vectors, while \(\mathbf{V}\) is an \(n\times n\) orthogonal matrix of the right-singular vectors, respectively.
The non-negative singular values are contained in the \(m\times n\) diagonal matrix \(\mathbf{\Sigma}\).  
The numerical rank of \(\tilde{\mathbf{J}}\) is determined from the number of non-negligible singular values in \(\mathbf{\Sigma}\). 
A full column rank indicates that all model parameters are locally identifiable. 
Conversely, rank deficiency implies the existence of a non-trivial null space, 
\begin{equation} 
    \mathcal N(\tilde{\mathbf J}) = 
    \left\{ 
        \Delta\mathbf p \in \mathbb{R}^n\,\middle|\,\tilde{\mathbf J}\Delta\mathbf p = \mathbf 0 
    \right\}, 
\end{equation}
whose dimension is given by 
\begin{equation}
    \dim\mathcal N(\tilde{\mathbf J}) =
    n - \operatorname{rank}(\tilde{\mathbf{J}}),
\end{equation}
where \(n\) is the number of model parameters. 
Parameter combinations contained in this null space do not affect the measurements and therefore cannot be uniquely identified from the available data. 
The corresponding right singular vectors associated with vanishing singular values directly reveal the unobservable combinations of stiffness tensor components. 
Beyond assessing parameter identifiability, the same framework can be used to evaluate and optimize measurement configurations. 
For example, the effect of reducing the number of propagation directions can be quantified by monitoring changes in the rank and null-space dimension of the normalized Jacobian. 
Likewise, the benefit of additional measurements can be assessed by augmenting the measurement vector and determining whether the associated null space is reduced. 
The analysis therefore provides a systematic means of identifying observable combinations of elastic constants, detecting redundant measurements, and assessing the value of complementary measurements in reducing parameter ambiguity.

In practice, singular values are rarely exactly zero due to numerical errors and finite-difference approximations. 
Therefore, the numerical rank is determined using a threshold on the normalized singular values, \[ \frac{\sigma_i}{\sigma_1} < \varepsilon, \] where \(\varepsilon\) is a prescribed tolerance. 
The corresponding condition number, \[ \kappa=\frac{\sigma_1}{\sigma_r}, \] with \(\sigma_r\) denoting the smallest retained singular value, provides an indication of how clearly identifiable and non-identifiable parameter combinations are separated.

\printbibliography

\end{document}